\documentclass[aps,prl,superscriptaddress,reprint]{revtex4-1}  

\usepackage{amsmath, amssymb}
\usepackage{xspace}
\usepackage{overpic}

\begin{document}

\title{Modular Linear Optical Circuits}

\author{Paolo L.~Mennea}
\affiliation{Optoelectronics Research Centre, University of Southampton, SO17 1BJ, United Kingdom}

\author{William R.~Clements}
\affiliation{Clarendon Laboratory, University of Oxford, Parks Rd., OX1 3PU, United Kingdom}

\author{Devin H.~Smith}
\affiliation{Optoelectronics Research Centre, University of Southampton, SO17 1BJ, United Kingdom}

\author{James C.~Gates}
\affiliation{Optoelectronics Research Centre, University of Southampton, SO17 1BJ, United Kingdom}

\author{Benjamin J.~Metcalf}
\affiliation{Clarendon Laboratory, University of Oxford, Parks Rd., OX1 3PU, United Kingdom}

\author{Rex H.S.~Bannerman}
\affiliation{Optoelectronics Research Centre, University of Southampton, SO17 1BJ, United Kingdom}

\author{Roel~Burgwal}
\affiliation{Clarendon Laboratory, University of Oxford, Parks Rd., OX1 3PU, United Kingdom}

\author{Jelmer J.~Renema}
\affiliation{Clarendon Laboratory, University of Oxford, Parks Rd., OX1 3PU, United Kingdom}

\author{W.~Steven Kolthammer}
\affiliation{Clarendon Laboratory, University of Oxford, Parks Rd., OX1 3PU, United Kingdom}
\affiliation{Department of Physics, Imperial College London, Prince Consort Rd., SW7 2BB, United Kingdom}

\author{Ian A.~Walmsley}
\affiliation{Clarendon Laboratory, University of Oxford, Parks Rd., OX1 3PU, United Kingdom}

\author{Peter G. R.~Smith}
\affiliation{Optoelectronics Research Centre, University of Southampton, SO17 1BJ, United Kingdom}

\date{\today}

\begin{abstract}
We propose and demonstrate a modular architecture for reconfigurable on-chip  linear-optical circuits. Each module contains 10 independent phase-controlled Mach-Zehnder interferometers; several such modules can be connected to each other to build large reconfigurable interferometers. With this architecture, large interferometers are easier to build and characterize than with traditional, bespoke, monolithic designs. We demonstrate our approach by fabricating three modules in the form of UV-written silica-on-silicon chips. We characterize these chips, connect them to each other, and  implement a wide range of linear optical transformations. We envisage that this architecture will enable many future experiments in quantum optics.
\end{abstract}

\maketitle

Integrated photonics is a promising platform for implementing the large-scale mode transformations required for optical quantum information processing.~\cite{OBrien09}. Photonic chips are more compact, more stable, and easier to scale up to large sizes than alternative platforms such as bulk optics. A wide range of quantum information protocols have thus been performed on-chip, using several different materials and architectures~\cite{Metcalf14,Harris17,Carolan15}. In addition, large integrated interferometric circuits have applications beyond quantum information science in telecommunication and classical computing~\cite{Capmany16,Shen17}. To make full use of these capabilities, the development of reconfigurable integrated devices that can be used for many applications is highly desirable. In the same way that programmable electronic chips have allowed for the development of modern computing, programmable optical chips are expected to play an important role in the development of quantum optics and photonics.

\begin{figure}[tb]
\begin{centering}
 \includegraphics[width=0.9\linewidth,trim={0.4cm 0 0.4cm 0},clip]{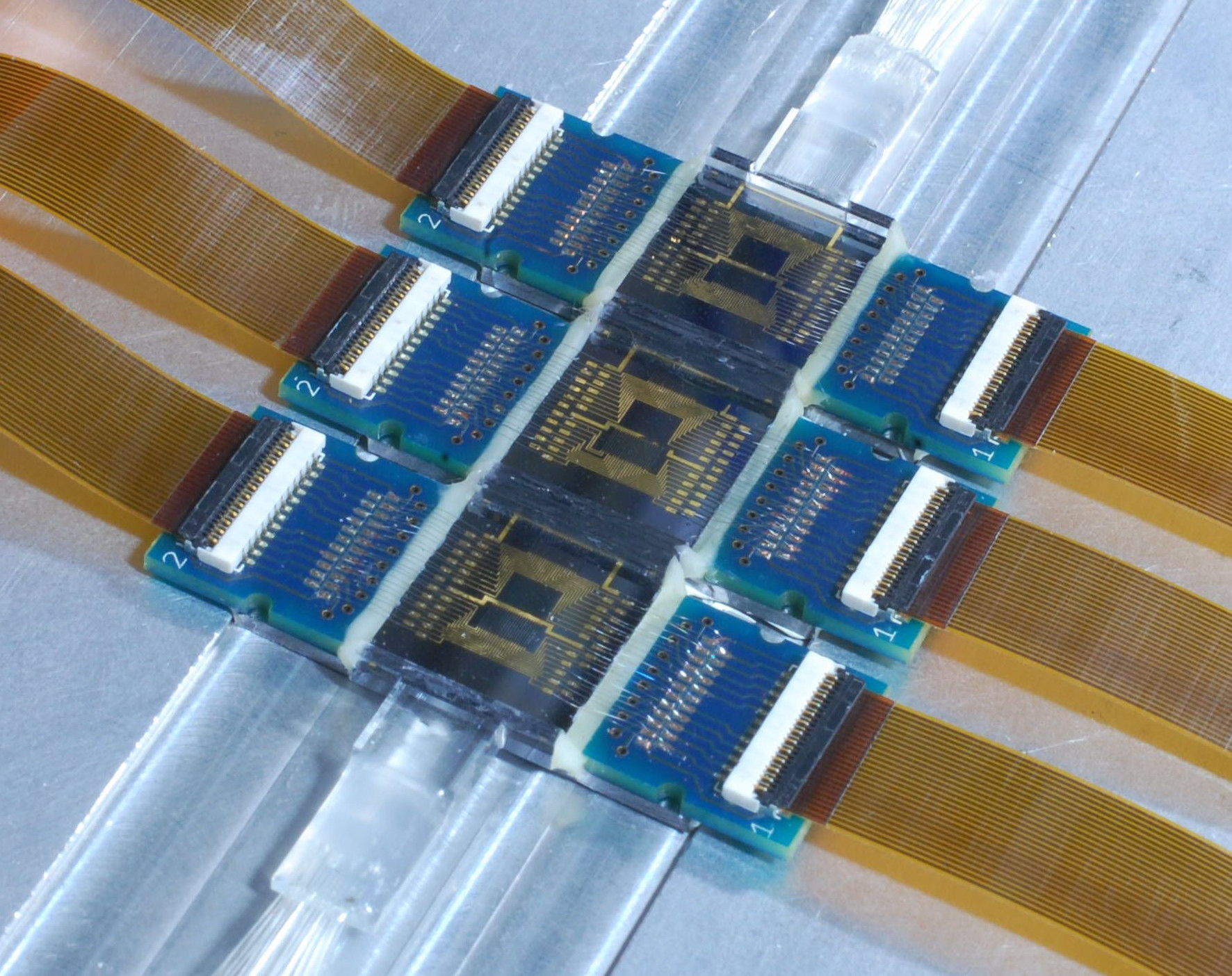}
 \end{centering}
 \caption{\label{fig:oneMZ} Three modules of our modular, reconfigurable quantum circuits are connected to build a large and programmable multiport interferometer with 20 inputs and outputs. This three module device is glued to two fiber V-groove assemblies and the phase shifters are controlled via electrical ribbon cables. The connectivity of the interferometer is determined by the number of modules.} 
\end{figure}

However, all the devices demonstrated to date are monolithic and are designed for a fixed number of optical modes. A chip designed to manipulate a large number of modes and complex interference may not be the most appropriate for also handling a smaller number of modes or simpler interference. Indeed, as device size increases to handle more modes, so do optical loss, optical crosstalk, and the complexity of characterizing all the optical elements on a chip. In addition, fabrication tolerances become more stringent~\cite{Burgwal17,Russell17}, since a single faulty component may jeopardize the correct operation of the entire device.

In this work, we propose and demonstrate the use of identical, flexible building blocks to compose interferometric circuits of any size. Each of these blocks consists of a column of Mach-Zehnder interferometers (MZI) with phase shifters both internally and on the input arms, as shown in fig.~\ref{fig:oneMZ}. Figure~\ref{fig:array} illustrates how different modules may be connected to construct a large interferometric circuit. Each of these modules may be tested and characterized individually, and imperfections in the modules can be mitigated by selecting those most suitable for a given experiment. Individual chips can be added or removed depending on the desired application.

In the following sections, we present our work on designing, characterizing, and assembling modular devices for use in quantum optics experiments. We first introduce the general principles of our modular approach, and discuss design considerations for the optical components on these chips. We then describe their fabrication method and our characterization results. Finally, we assemble a three-chip device and demonstrate its use by implementing a wide range of optical transformations.

\section{Modular Design}

\begin{figure}[tb]
\centering
 \begin{overpic}[width=\linewidth]{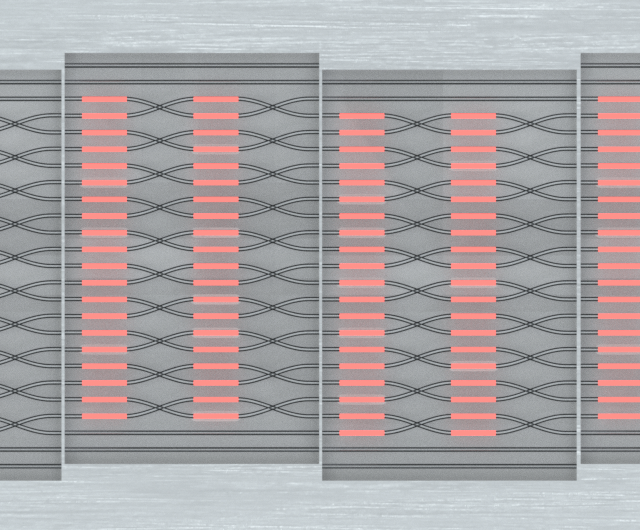} 
 \put (10,9) {\huge$\displaystyle\underbrace{\hspace{3.5cm}}_\text{Module}$}
 \end{overpic}
 \caption{A\label{fig:array} rendered impression of concatenated modular chips. The phase shifters are shown in pink, and the waveguide pitch has been exaggerated for clarity. Note that each second chip is offset by one waveguide (127$\mu$m) to form a multiport interferometer.} 
\end{figure}

Our module consists of 10 tunable MZIs placed side by side. These modules are intended to be tiled to build larger interferometers. For example, as shown by Clements \textit{et al.}~\cite{Clements16}, $N$ such modules each at least $N/2$ MZI wide suffice to perform an arbitrary unitary operation on $N$ modes, with optimal circuit depth.  Here, each MZI acts as an arbitrarily reconfigurable beamsplitter, with two phase-shifters acting as the tuning elements: one internal phase setting the reflectivity and one external phase (either on the input or output) setting the phase between elements.  The modules are then interlaced, as shown in fig.~\ref{fig:array}, with the left-hand outputs of one MZI fed into the right-hand inputs of the next layer of MZIs to form the $N$ mode unitary operation with $N$ modules.

A modular approach gives us several advantages over monolithic designs. 
First, it is easier in practice to work with individual layers of MZIs than it is to work with a large interferometer. Optical characterization is simpler when the modes can be addressed in small, discrete, groups. Optimizing the fabrication of a single layer of MZIs is also easier than optimizing that of an entire interferometer. Furthermore, faulty components can be replaced without compromising the entire structure.

Second, the modular approach provides experimental flexibility due to reconfigurability. Many  protocols in linear optics---such as quantum teleportation \cite{Metcalf14}---do not require fully connected interferometers. For these applications, additional layers of MZIs are not necessary and merely increase loss. On the other hand, as discussed by Burgwal \textit{et al.} \cite{Burgwal17}, additional MZIs at the output of a universal multiport interferometer can be used to increase the fidelity of the overall transformation. Our modular approach allows for the addition of these MZIs if necessary. Alternatively, our approach also allows us to implement the nested-MZI architecture proposed by Miller \cite{Miller15} which, at the cost of additional beamsplitters, allows for much higher fabrication error. 

The downsides of using a modular approach are that additional coupling steps are required, and that additional loss can occur in two ways: first, there is some amount of interface loss between modules; and second, the total length of waveguide in the device is unavoidably lengthened to allow for the waveguides to match at the interfaces. However, chip to chip coupling can be automated. Furthermore, as the waveguides are fabricated using the same process, the chips are very well mode-matched. As we show later, in combination with index-matched adhesive we can achieve very low coupling losses.

\section{Component Design}

\subsection{Choice of platform}

We choose UV-written silica on silicon as the platform for our modular chips. This choice is motivated by the exceptionally low coupling loss to fiber and propagation loss that can be achieved with this platform \cite{Zauner98}. They are therefore compatible with many state of the art photon sources and detectors that are designed to be fiber coupled. Furthermore, losses are a significant concern for quantum optics experiments and must be minimized. Moreover, UV writing does not require a lithographic step, which allows for quick turnaround times from planning to production.

\subsection{X-couplers}

We use X-couplers~\cite{Kundys09} instead of the more typical directional couplers. These X-couplers can be thought of as the zeroth-order version of a directional coupler---incoming light couples evanescently between the modes for less than one-quarter of a sine wave while transiting the device.  Using these couplers instead of directional couplers has several effects: first, the bandwidth is maximized for an evanescent device; second, the coupler has a compact footprint due to the small coupling region; and third,  the device properties are defined primarily by the crossing angle of the two guides, improving fabrication tolerances.

\subsection{Thermo-optic phase-shifters}

The final ingredient necessary to fabricate a fully tunable device using an MZI is two phase-shifters: one inside the MZI, and one on either the input or output ports.  The use of microheaters as such a phase shifter is a tried-and-tested approach in integrated quantum optics~\cite{Smith09,Carolan15}. Such phase shifters have high stability and tuning range but a slow response on the order of milliseconds.  Nonetheless, this response allows the device to be configured as required and remain at a fixed point during operation, which is sufficient for many protocols of interest.  

In order to improve stability, \emph{both} modes in each MZI have a phase shifter attached (see fig.~\ref{fig:array}), and the total amount of current passed through the two phase-shifters is held constant, allowing for push-pull operation of the phase: this increases the tuning range per unit length by a factor of two, and also ensures that the total amount of heat dissipated in each portion of the chip is constant, greatly reducing cross-talk and improving stability.  

\section{Fabrication}

The silica glass layers are fabricated in-house by FHD on a silicon wafer; a 15$\mu$m thermal oxide layer forms the undercladding, onto which is deposited a 4$\mu$m core layer doped with germanium and boron to promote photosensitivity. This is followed by an 8$\mu$m boron and phosphorus doped upper cladding layer. Dopant levels in  core and cladding are adjusted to match the refractive index profile to the thermal oxide.

Waveguides are directly written using a 244~nm frequency-doubled Ar:Ion laser into this planar structure: a focused UV beam is translated relative to the photosensitive sample on precision air-bearing stages, producing buried channel waveguides. This technique allows waveguides with relatively large modes well matched to optical fiber to be produced, minimizing coupling losses while reducing the required alignment tolerances between modules. A key advantage of this technique is that it permits the inscription of programmatically-controlled first-order Bragg gratings during the waveguide writing process, which may be written out-of-band to aid classical characterization of each module.

To enhance the refractive index change the devices are kept in a high-pressure hydrogen atmosphere for several days; however due to outgassing of the hydrogen the maximum writing time per chip is limited to about an hour. Thus, for time-limited fabrication techniques like ours a further benefit of the modular architecture is that the individual modules can be written serially, greatly increasing the number of possible devices fabricable in our laboratory.

The phase-shifters are patterned through contact lithography and lift-off of a 170~nm e-beam evaporated nichrome layer, while the wiring is a 200~nm copper layer deposited in the same manner. Computer control of the on-chip phase shifters was accomplished using custom drive electronics, producing an array of pulse-width-modulated drive signals at up to 20~V with 8-bit resolution.

\section{Characterization}

\begin{figure*}[tb]
\centering
\includegraphics[width=0.9\textwidth]{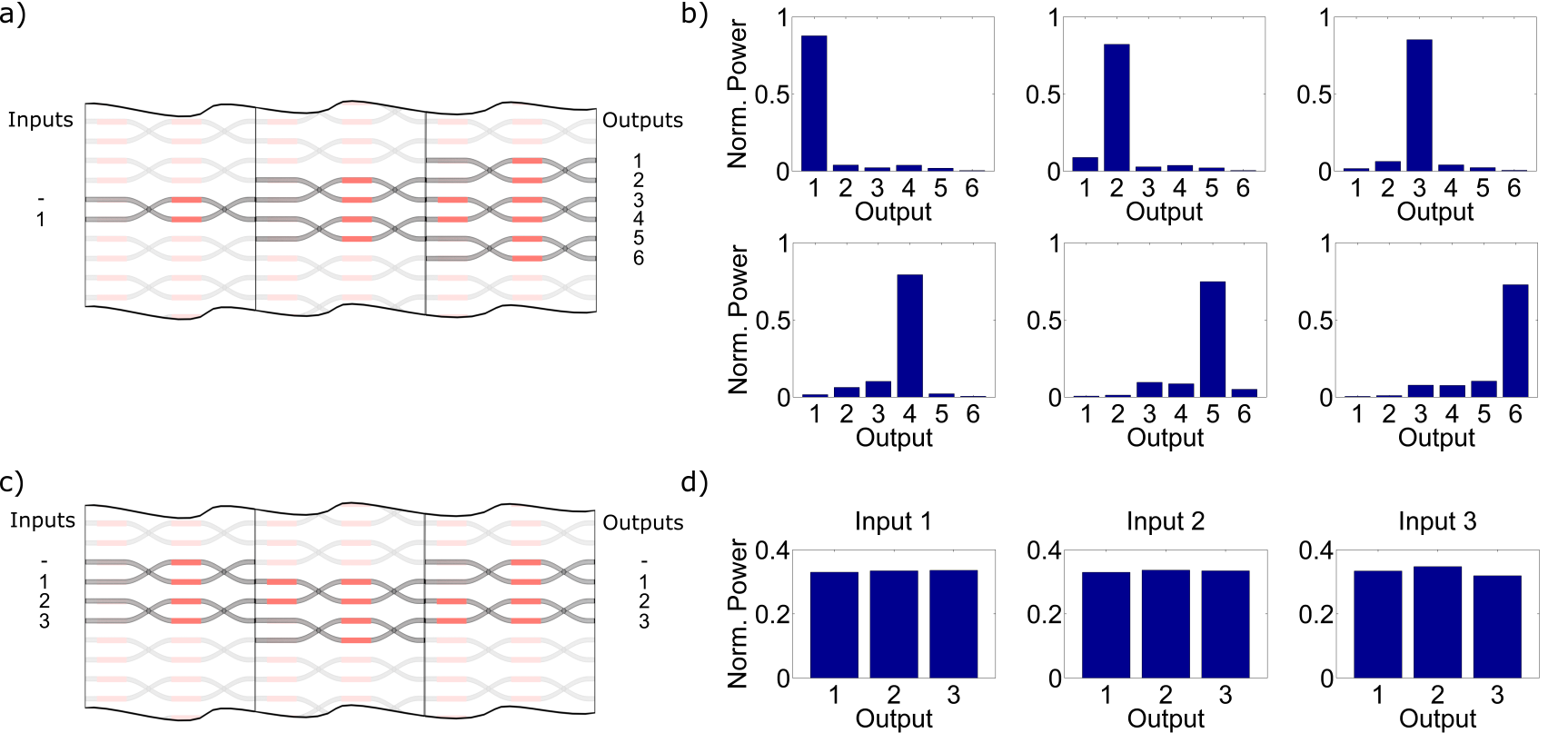}
 \caption{\label{fig:data} Experimental results demonstrating the programmability of our three chip device. a) Our device can be configured to act as a $1{\times} 6$ switch. The six relevant phase shifter pairs that we use to program the switch are schematically represented by red bars in the circuit. 
b) Bar plots of the six output intensities when the device is configured to switch the light towards a given output, for all six possible outputs. c) Our device can also be configured to act as a balanced $3 \times 3$ interferometer (a tritter) using the four phase shifter pairs shown in this circuit. d) Bar plots of the three output intensities for all three inputs when the device is configured as a tritter.} 
\end{figure*}
After fabrication, the individual modules can be characterized before assembly: both the fixed parameters set by the particular fabrication run as well as the tuning parameters of the phase-shifters must be found. This characterization process is easier for our modular devices than for monolithic circuits since each MZI can be addressed individually. The emission from the two output ports of each MZI, as well as reflectance data from the Bragg gratings embedded in the waveguides before, inside, and after the MZI, yield the parameters of interest: the splitting ratios, excess loss of the couplers, an estimate of the facet loss, and the phase-shifters' tuning curves. The phase-shift of the other pair of couplers can also be found either at this stage by inputting coherent light on both inputs of an MZI or determined after the modules are assembled.   
\begin{table}[t]
\caption{\label{tbl:char}
Typical measured parameters of interest for fabricated modular chips, as determined from single-module characterization, at an operating wavelength of 780~nm.}
\begin{tabular}{lr@{}l}
   Fiber coupling loss & 0.3 & $\pm$0.1 dB\\	
   Chip-chip facet loss & 0.2 & $\pm$0.1 dB\\
   Propagation loss & 0.35& $\pm$ 0.04 dB/cm \\
   Phase tuning range & (2.7 &  $\pm$0.2) $\pi$ rad \\
   Coupling ratio & 57& $\pm$4\% \\
   Coupler excess loss & 2.1&  $\pm$ 0.3 dB \\
      \end{tabular}
\end{table}
Table \ref{tbl:char} shows  typical values for the parameters of interest measured this way for the chips used here. Fiber to chip coupling losses are similar to what has been demonstrated with recent on-chip quantum experiments~\cite{Carolan15}, but could still be improved. Chip to chip coupling losses are low, as expected. However, the coupler excess loss is particularly high, and the splitting ratios are quite far from the 50\% that is required to construct fully tunable MZIs. This is a fabrication issue that will be corrected in future work. 

We also find that our thermal crosstalk measurements justify the use of our dual heater design. To measure crosstalk, a MZI is placed close to its 50:50 point and its outputs measured as settings of adjacent heaters were varied. This is carried out both in the case of complementary heating and with one heater in each pair disabled. We find that when achieving a $\pi$ phase shift on the target MZI, our dual heaters induce a crosstalk of about 0.01$\pi$ to the neighboring MZI and of 0.007$\pi$ to the next-nearest MZI, whereas using a single phase shifter to achieve the same phase shift induces about twice as much crosstalk.

\section{Experimental Results}
\label{sec:exp}

To perform a demonstration of these devices, we assemble three of the modular chips, as shown in Fig. \ref{fig:oneMZ}. Three chips is the minimal number that we can use to demonstrate a $3 \times 3$ universal multiport interferometer: a device that can realize any linear transformation between the input and output optical channels. Furthermore, for future applications in linear optics, three chips is the number that is required to perform quantum teleportation \cite{Metcalf14}, which is an important building block of linear optical quantum information schemes \cite{Knill01}. We show that our assembly can implement a wide range of useful optical transformations.

As a first demonstration of our three chip assembly, we show that light from one input can be switched to any of its six available outputs. The procedure for doing so is straightforward. We send light into the input, monitor the output that we seek to switch the light into, and sequentially optimize the MZIs on each chip along the path to that output to maximize the measured power. Our results are shown in Fig.~\ref{fig:data}b. We see that our chip assembly successfully routes most of the light to the desired output, for all six outputs. However, some leakage does occur, which may originate from both the imperfect splitting ratios of the beam splitters and some light recoupling into the waveguides after being lost.

Next, we show that our chip assembly can implement a balanced $3 \times 3$ interferometer, also known as a tritter. Tritters can be used for fundamental studies of quantum interference \cite{Menssen17}, as well as for classical photonics applications. We use a self-configuring approach to implement a tritter \cite{Miller15}, which does not require detailed prior characterization of the phase shifters. Our results are shown in Fig. \ref{fig:data}d. We see that the light is indeed equally split between all output ports.

More generally, we also demonstrate that our chip assembly implements a $3 \times 3$ universal multiport interferometer. We follow the the algorithimic method of Clements \textit{et al} \cite{Clements16}. We first fully characterize the $2 \times 2$ transformations implemented by the relevant MZIs and phase shifter for all of their phase shifter settings using an automated procedure. We then randomly select a $3 \times 3$ unitary matrix $U$ to be implemented with our device. We perform a decomposition procedure on $U$ that yields the correct phase shifter settings \cite{Clements16}, which we program into our device. To test the performance of this protocol, we  then measure the $3 \times 3$ transfer matrices corresponding to all 9 input-output relations, which we normalize to remove the effect of loss. We repeat this process for 50 randomly selected unitary matrices and experimentally measure the transfer matrices for all of them. We compare these transfer matrices to the targeted transformations and find an average transformation fidelity of 97.5\%, which shows that our implementation of a universal multiport interferometer is successful. 

\section{Conclusion}

We have designed and fabricated a modular system for implementing interferometers of arbitrary size. This system is easier to characterize than a monolithic interferometer, as individual elements can be interrogated, and a single out-of-spec component can simply be replaced. In addition, our particular implementation of the modular system is fabricated with fiber-compatible silica waveguides, reducing loss both at the internal and external interfaces to quantum-compatible tolerances; we additionally used push-pull phase-shifters, which substantially reduce crosstalk between neighboring devices. Going forward, these devices will be used to implement a wide range of quantum interferometric protocols.

\bibliography{modular}
\end{document}